\documentclass[preprint,aps,prd,showpacs,groupedaddress,floatfix,nofootinbib]{revtex4}

\usepackage{dcolumn}% Align table columns on decimal point
\usepackage{graphicx}
\usepackage{epsfig}
\usepackage{hyperref}
\usepackage{graphicx}% Include figure files
\usepackage{dcolumn}% Align table columns on decimal point
\usepackage{amssymb,eucal}
\usepackage{longtable}
\usepackage{amsmath,amssymb,bm}
\usepackage{mathrsfs}

\newcommand{\mathd}{\mathrm{d}}

\newcommand{\mathe}{\mathrm{e}}

\newcommand\Tr{{\rm Tr }}

\def\Tr{\mathop{\rm Tr}\nolimits}

\begin{document}
%\preprint{\vbox{\hbox{BARI-TH-2013-6?? \hfill}}}
\title{Thermal power terms in the Einstein-dilaton system}
\author{Fen Zuo}
\affiliation{
School of Physics, Huazhong University of Science and Technology, Wuhan 430074, China
}

\begin{abstract}
We employ the gauge/string duality to study the thermal power terms of various thermodynamic quantities in gauge theories and the renormalized Polyakov loop above the deconfinement phase transition. We restrict ourselves to the five-dimensional Einstein gravity coupled to a single scalar, the dilaton. The asymptotic solutions of the system for a general dilaton potential are employed to study the power contributions of various quantities. If the dilaton is dual to the dimension-4 operator $\Tr F_{\mu\nu}^2$, no power corrections would be generated. Then the thermal quantities approach their asymptotic values much more quickly than those observed in lattice simulation. When the dimension of the dual operator is different from $4$, various power terms are generated. The lowest power contributions to the thermal quantities are always quadratic in the dilaton, while that of the Polyakov loop is linear.
%The asymptotic expansions are then extrapolated to the medial temperature region. With such an extrapolation
As a result, the quadratic terms in inverse temperature for both the trace anomaly and the Polyakov loop, observed in lattice simulation, cannot be implemented consistently in the system. This is in accordance with the field theory expectation, where no gauge-invariant operator can accommodate such contributions.
%Therefore, these quadratic contributions should be due to some additional fields in the bulk.
Two simple models, where the dilaton is dual to operators with different dimensions, are studied in detail to clarify the conclusion.
%We conclude that the quadratic thermal contributions can not be generated consistently in the Einstein-dilaton system.

\end{abstract}

%\keywords{Gauge/string duality; Skyrmions; 1/Nc expansions.}
\pacs{
11.25.Tq, %Gauge/string duality
%12.39.Dc, %Skyrmions
11.15.Pg %Expansions for large numbers of components (e.g., 1/Nc expansions)
%11.10.Kk, %Field theories in dimensions other than four
%11.15.Tk  %Other nonperturbative techniques
%12.38.Lg  %Other nonperturbative calculations
}

 \maketitle

\section{Introduction}
Lattice simulations  have disclosed  interesting properties of the gauge theory thermodynamics.
In \cite{Meisinger:2001cq,Pisarski:2006yk} it is shown that the lattice data  for the trace anomaly in  pure SU(3) gauge theory \cite{Boyd:1996bx} are dominated by a $T^2$ term in the temperature region $1.1~T_c<T<4T_c$. From this observation, a ``fuzzy bag" model has been advertised, which allows to express the pressure $p_{\rm{QCD}}(T)$ as~\cite{Pisarski:2006yk}
\begin{equation}
p_{\rm{QCD}}(T)\approx f_{\rm{pert}} T^4-B_{\rm{fuzzy}}T^2-B_{\rm{MIT}}+...
\end{equation}
Lattice analyses also show that the constant term is small, $B_{\rm{MIT}}\ll B_{\rm{fuzzy}}T_c^2$. Therefore, one expects $B_{\rm{fuzzy}}\approx f_{\rm{pert}} T_c^2$ to ensure $p(T_c)\approx ~0$ at the critical temperature $T_c$ for the pure glue theory. From the  expression for the pressure one  derives the  trace anomaly $\theta(T)$,
\begin{equation}
\theta(T)\approx 2B_{\rm{fuzzy}}T^2+4B_{\rm{MIT}}+...
\end{equation}
Further investigations  show that, for various $N_c$,  the trace anomaly in SU($N_c$) gauge theory  can be  represented  by the formula  \cite{Panero:2009tv}
\begin{equation}
\frac{\theta(T)}{T^4} = \frac{\pi^2}{45} (N_C^2-1) \cdot \left(
1 - \frac{1}{\left\{
1 + \exp \left[ \frac{(T/T_c)-f_1}{f_2}
\right] \right\}^2}
\right) \left( f_3\frac{T_c^2}{T^2} + f_4\frac{T_c^4}{T^4}
\right).
\label{eq.theta0}
\end{equation}
As $f_2$ turns out to be  tiny and $f_1$ smaller than $1$, this formula simplifies above $T_c$:
\begin{equation}
\theta(T) = \frac{\pi^2}{45} (N_C^2-1)   \left( f_3 T_c^2~ T^2 + f_4 T_c^4\right) ,
\label{eq.theta1}
\end{equation}
with the large-$N_C$ extrapolation values $f_3\simeq 1.768$ and $f_4\simeq -0.244$.
%It indicates $\theta$ is nearly a linear %function of $T^2$, with a small constant correction.
Since in the free field limit one has $f_{\rm{pert}}\simeq  \frac{\pi^2}{45} (N_C^2-1)$, this gives $B_{\rm{fuzzy}}\simeq 0.884~f_{\rm{pert}} T_c^2$, and $B_{\rm{MIT}}\simeq -0.061 ~f_{\rm{pert}} T_c^4$:
hence,  the  expectation for the pure gauge theory is roughly met. Notice, however, that the bag constant in pure gauge theory turns to be negative, in contrast to that in QCD at zero temperature and finite temperature \cite{Bazavov:2009zn}. Recent lattice simulations also show that the trace anomaly in 3D SU(N) gauge theory is dominated by a $T^2$ term in the high temperature region~\cite{Bialas:2008rk,Caselle:2011mn}; this indicates a large linear contribution in inverse temperature, in contrast to the quadratic contribution in 4D. Further lattice calculation for the bulk viscosity in \cite{Meyer:2007dy} also indicates that it can be described by a quadratic term in inverse temperature above the phase transition~\cite{Rajagopal:2009yw}.

In Ref.~\cite{Megias:2005ve} it was found that the lattice results in \cite{Kaczmarek:2002mc} for the logarithm of the Polyakov loop $L(T)$ above the phase transition, in SU(3) pure gauge theory,  can be fitted by a linear function of  $1/T^2$:
\begin{equation}
-2\log\left[ L(T)\right]\simeq  a+b\left(
\frac{T_c}{T}\right)^2.\label{eq.logL}
\end{equation}
%with $a=-0.27(5),~b=1.81(13)$ for the lattice data with $N_\tau=4$, and $a=-0.23(1),~b=1.72(5)$ for %$N_\tau=8$~\cite{Kaczmarek:2002mc}.
A similar pattern is also obtained in lattice simulations for SU(4) and SU(5) pure gauge theory ~\cite{Mykkanen:2012ri}. %with the fitted parameters $a=-0.0959(11), ~b=1.1166(55)$ for $N_C=4$, and $a=-0.3056(15),~b=1.4283(62)$ for %$N_C=5$~\cite{Mykkanen:2012ri}.
Notice that, although the Polyakov loop is renormalization scheme-dependent, it is argued that the linear correction vanishes in the phenomenology preferred scheme. To eliminate the scheme dependence, it is more convenient to study the heavy quark free energy,  related to the Polykov loop as
\begin{equation}
L(T)=\mathe^{-F_Q(T)/T}.
\end{equation}
With the simple form (\ref{eq.logL}),  the derivative of $F_Q$ would be
\begin{equation}
\frac{\mathd F_Q(T)}{\mathd T}\simeq \frac{a}{2} -\frac{b}{2}\left(
\frac{T_c}{T}\right)^2. \label{eq.dFdT2}
\end{equation}
Similar to the fuzzy bag model, one may estimate  the parameters $a$ and $b$, on the basis of the properties of the renormalized Polyakov loop at $T_c$ and in the high temperature limit. Perturbation theory predicts that,  in the high temperature limit,  the renormalized Polyakov loop should approach one from above. This is confirmed by  lattice simulations~\cite{Kaczmarek:2002mc,Mykkanen:2012ri}, which agree quite well with the perturbative results at large temperature. Subtracting the perturbative contributions from the lattice data, one would expect that the remaining contribution with the form (\ref{eq.logL}) should give $a\approx0$. Moreover, both for SU(4) and SU(5)  the lattice data show that $L$ approaches $0.5$ when $T\to T_c^+$~\cite{Mykkanen:2012ri}. If the dominance of the quadratic term can be valid near the phase transition, $b$ is fixed to $b\approx 2 \log 2\simeq 1.386$. These results could be interpreted of as the nonperturbave values of $a$ and $b$ in the large-$N$ limit, which are consistent with the results at $N_C=3,4,5$~\cite{Megias:2005ve,Mykkanen:2012ri}. However, such an estimate may not be quite reliable, since higher power terms may become important close to $T_c$.

%Since the derivative of $F_Q$ is unchanged when $F_Q$ is shifted by a constant, it will be easier to check this %expression at the first stage.

 %It is found in \cite{Megias:2005ve} that, the logarithmic of the renormailzed Polyakov loop in SU(3) gauge theory, exhibits unequivocal quadratic contributions inverse temperature squared above the phase transition. Recently such a contribution is also observed for SU(4) and SU(5) pure gauge theory~\cite{Mykkanen:2012ri}. Although the Polyakov loop is renormalization scheme-dependent, in the phenomenology preferred scheme the linear correction vanishes and the quadratic term dominates.

 %The leading constant term turns to be small and consistent with perturbative prediction at large temperature, while the %quadratic term signals some significant nonperturbative contribution.

 %A similar pattern is observed in \cite{Pisarski:2006yk} for the the trace anomaly. In the region $1.1~T_c\leq T \leq 4T_c$, the early lattice data \cite{Boyd:1996bx} for the trace anomaly in SU(3) gauge theory can be well described by a single $T^2$ term. This temperature dependence of the trace anomaly continues for SU($N_c$) gauge theory with $N_c=4,5,6,8$~\cite{Panero:2009tv}. Recent lattice simulation also shows that the trace anomaly in 3D SU(N) gauge theory is dominated by a $T^2$ term in a large temperature region~\cite{Caselle:2011mn}. This indicates a large linear contribution in inverse temperature, in contrast to the quadratic contribution in 4D.

Many explanations have been proposed for the quadratic contributions. The corresponding $T^2$ term in the pressure can be viewed as a temperature-dependent correction to the constant bag term. This is accommodated in the fuzzy bag model by generalizing the usual bag term to a series in powers of inverse temperature squared~\cite{Pisarski:2006yk}. In the high temperature region, the $T^2$ term could be considered as the correction to the leading perturbative prediction. This is  similar to the case of heavy quark potential at zero temperature~\cite{Akhoury:1997by}. A natural way to generate such a contribution consists in introducing a  dimension-two gluon condensate~\cite{Megias:2005ve,Megias:2009mp}; however, since no gauge-invariant operator of dimension two exists, the properties of such a condensate cannot be derived systematically.

At large $N$, the thermodynamics of SU(N) gauge theory can be  described by the black hole solutions on the gravity side, according to the gauge/string duality~\cite{Maldacena:1997re,Gubser:1998bc,Witten:1998qj,Witten:1998zw}. For ${\mathcal N}=4$ Super Yang-Mills theory in flat spacetime, the thermodynamic quantities at strong coupling are found to be exactly $3/4$ of the free coupling results~\cite{Gubser:1996de}. It is thus concluded that thermodynamics is insensitive to the coupling~\cite{CasalderreySolana:2011us}. However, the novel temperature dependence of various quantities, shown above, must have a nonperturbative origin. According to the duality dictionary~\cite{Gubser:1998bc,Witten:1998qj},  a two-dimensional condensate would correspond to the quadratic gravitational fluctuations around asymptotic Anti-de Sitter~(AdS) background. At finite temperature, such fluctuations are naturally generated when the boundary of the AdS spacetime is taken  $S^1\times S^3$ instead of $S^1\times R^3$~\cite{Hawking:1982dh,Witten:1998zw}. Indeed, quadratic terms appear in all the thermodynamic quantities, and also in the renormalized Polyakov loop~\cite{Zuo:2014vga}. As an analog of the MIT bag model, the hard-wall model \cite{deTeramond:2005su,Erlich:2005qh,DaRold:2005zs} does not accommodate such condensates. In contrast, in the soft-wall model~\cite{Karch:2006pv,Andreev:2006vy} these contributions appear in a similar pattern as the fuzzy bags~\cite{Andreev:2007zv,Andreev:2009zk}. Such contributions continues to appear even when nonzero chemical potential is introduced~\cite{Colangelo:2013ila}. Unfortunately, in order to fit the thermodynamic quantities and the Polyakov loop, different background fluctuations are required. The soft-wall model can be consistently realized in the gravity-dilayon system~\cite{Csaki:2006ji,Gursoy:2007cb,Gursoy:2007er}. With proper choices for the dilaton potential, such a construction also exhibits the Hawking-Page phase transition as in pure AdS~\cite{Gursoy:2008bu,Gursoy:2008za}. By fine-tuning the parameters in the dilaton potential, the lattice data for various thermal quantities and transport coefficients can be reproduced quite accurately~\cite{Gubser:2008ny,Gursoy:2008bu,Gubser:2008yx,Gursoy:2008za,Gursoy:2009jd,Gursoy:2010fj}. It is natural to wonder whether and how the quadratic contributions found in lattice simulations are generated in such a system. However, due to the complexity of the potential, full analytic results are unavailable and numerical techniques have to be used. From the numerical results it is not easy to separate the different contributions.

%{Gubser:2008ny,Gursoy:2008bu,Gursoy:2008za}. It has also been extended to describe the behavior of the bulk %viscosity~\cite{Gubser:2008yx,Gursoy:2010fj}
%some modifications to the original model are needed. Since the model is not derived from a gravity system, such modifications are not within control.
%it is induced naturally from the infrared deformation. Indeed in such a model
%the quadratic contribution to both the trace anomaly and the Polyakov can be generated~\cite{Andreev:2007zv,Andreev:2009zk}.
%While in the hard-wall model \cite{Erlich:2005qh,DaRold:2005zs} no such condensate appears

An important observation  is that the thermal quantities are well approximated by the leading conformal term plus the quadratic term~\cite{Andreev:2007zv,Andreev:2009zk,Zuo:2014vga} . This is also the situation in the lattice results~\cite{Pisarski:2006yk,Panero:2009tv,Megias:2005ve,Mykkanen:2012ri}, where the deviations from such a truncated expansion are found only close to $T_c$. The reason is due to the absence of odd power contributions and the smallness of the quartic correction.
%Such an observation indicates that the truncated asymptotic expansion in the ultraviolet could be approximately valid %even in the medium temperature region.
Therefore, one could use such an expansion to investigate the quadratic contributions in the gravity-dilaton system. As shown in ~\cite{Hohler:2009tv,Cherman:2009tw}, the asymptotic solutions can be obtained analytically order by order, and even extended to arbitrary spacetime dimensions~\cite{Yarom:2009mw}. The asymptotic properties of various quantities, like the speed of sound~\cite{Hohler:2009tv,Cherman:2009tw}, the bulk viscosity~\cite{Cherman:2009kf,Yarom:2009mw} and the renormalized Polyakov loop~\cite{Noronha:2010hb}, have been studied in detail. These asymptotic results could be polluted, since in the high temperature region the perturbative contributions dominate. It would be more reasonable to study the power contributions in the intermediate temperature region with these asymptotic solutions, which we do now.

The paper is organized as follows. In the next Section we briefly describe the general form of the perturbative corrections. The general pattern of the power contributions in different thermal quantities are shown in Sect.III. In Section IV we give two simple models to illustrate explicitely the power contributions in different quantities. In the last Section a short discussion is given.

\section{Perturbative corrections}
Let us first consider the perturbative corrections.  The perturbative running of the coupling constant translates into the logarithmic behavior of the 't Hooft coupling in the ultraviolet~($z\to 0$)~\cite{Gursoy:2007cb,Gursoy:2007er},
\begin{equation}
\lambda\sim (-\log z)^{-1},
\end{equation}
in the conformal coordinate $z$. In the high temperature region, the background is approximately AdS-Schwarzschild black hole, with  temperature %the black hole horizon will be related to the temperature as
\begin{equation}
T\sim\frac{1}{\pi z_H},
\end{equation}
$z_H$ being the the black hole horizon.
Hence, the perturbative $\lambda$ series of the thermal quantities, in the ultraviolet, translates into a series of $(\log T)^{-1}$~\cite{Alanen:2009xs,Megias:2010ku}. This is expected from the operator product expansion in the boundary field theory~\cite{Pisarski:2006yk}. Since the logarithmic terms dominate over the power ones for extremely large $T$, the later can only be seen when those perturbative terms are turned off. However, it could not be excluded that the power terms may appear when an infinite series of perturbative terms are summed up~\cite{Narison:2009ag}. Also in the gravity approximation, asymptotic freedom and the logarithmic running are difficult to implement. Therefore, here  we  neglect these perturbative contributions, and focus on the power terms generated directly from some nonperturbative mechanism.

\section{General pattern for the thermal power terms}
\subsection{Asymptotic solutions}
Consider the action of gravity coupled to the dilaton with an arbitrary potential
%The general Einstein-dilaton action is given by
\begin{equation}\label{eq:S5}
S_5 = \frac{1}{2 \kappa^2} \left[\int_M \!\! d^5x \; \sqrt{-g}\, \left (R - V(\phi) - \frac{1}{2} (\partial \phi)^2 \right) - 2 \int_{\partial M} \!\!\!\! d^4x \; \sqrt{-\gamma}\, K \right],
\end{equation}
where $R$ is the Ricci scalar, $g$  the determinant of the metric,
$\gamma$  the determinant of the induced metric on the UV boundary
$\partial M$, $K$  the extrinsic curvature on $\partial M$, and $\kappa^2$ is related to the 5D Einstein gravitational constant,  $\kappa^2=8\pi G_5$. Since we do not consider perturbative running of the coupling, the dilaton potential has the following expansion near the boundary
\begin{equation}
\label{asympV}
V(\phi) = -\frac{12}{l^2} + \frac{m^2}{2}  \phi^2 + \mathcal{O}(\phi^{4}) \;.
\end{equation}
 The squared mass is related to the scaling dimension $\Delta_+$ of the gauge theory operator $\mathcal{O}_{\phi}$,   $m^2=\Delta_+(\Delta_+-4)/l^2$. The dilaton field goes to zero at the boundary and one recovers the asymptotic AdS spacetime with radius $l$.

To study the asymptotic solutions in the ultraviolet, it is convenient to work with the following ansatz~\cite{Hohler:2009tv,Cherman:2009tw}:
\begin{equation} \label{eq:ansatz1}
ds^2 = \frac{1}{z^2}\left(-f(z) dt^2 + d\vec{x}^2\right) + e^{2 B(z)} \frac{dz^2}{z^2 f(z)}.
\end{equation}
The equations of motion are:
\begin{eqnarray}
&\dot{B} = -\frac{1}{6}\, \dot{\phi}^2, \label{eq:ein1}\\
&\ddot{f} = \left({4} + \dot{B}\right) \dot{f}, \label{eq:ein2}\\
&-6 \dot{f} + f \left({24} - \dot{\phi}^2\right) + 2 {e^{2 B}}\, V(\phi) = 0, \label{eq:ein3}\\
&\ddot{\phi}f + \dot{\phi}\left(\dot{f} - f(4 + \dot{B})\right) - e^{2 B}\, V^\prime(\phi) = 0, \label{eq:ein4}
\end{eqnarray}
where a dot denotes a $\log z$ derivative (e.g., $\dot\phi=z\,d\phi/dz$), while $V'=dV/d\phi$.
For high temperature, the black hole horizon is close to the boundary, the dilaton is small everywhere and the gravity-dilaton system can be solved recursively in $\phi$. Moreover, from Eqs.(\ref{eq:ein1})-(\ref{eq:ein4}) one can see that the dilaton backreacts on the the metric functions $B(z)$ and $f(z)$ at even orders in $\phi$, while the metric functions backreact on the dilaton at odd orders in $\phi$, and this  is a  simplification for the recursion procedure. Explicit solutions of the system with $\Delta_+=1/2,1,2$ are reported  in Ref.~\cite{Li:2011hp}, from which one can check the above assertions.

Since the equations for $\phi$ and $f$ are of second order, while that for $B$ is of first order, there are in principle five integration constants. Lorentz invariance at the boundary requires $f(\epsilon)=1$, and the equation for $f$ can be integrated:
\begin{equation}
\dot{f}=-f_0 z^4 \mathe^B,
\end{equation}
with the constant $f_0$ depending on the horizon $z_H$. The boundary condition for $\phi$ should be
\begin{equation}
\phi(\epsilon)=c~ \epsilon^{\Delta_-}
\end{equation}
with $\Delta_-=4-\Delta_+$. Then, the other constant is determined by the regularity on the horizon. Finally, the function $B$ is completely determined by $B(\epsilon)=0$, which ensures that the spacetime is asymptotically AdS. Thus, the full solutions depend on two parameters, $c$ and $z_H$.

At zero order in $\phi$, the system is simply the AdS Schwarzschild black hole
\begin{equation}
B(z)=0,\quad  f(z)=1- \frac{z^4}{z_H^4}.
\end{equation}
At first order, the dilaton is solved~\cite{Hohler:2009tv,Cherman:2009tw}
\begin{equation}
\phi(z) = \phi_H \, {}_2F_1\left(1-\frac{\Delta_+}{4},\,\frac{\Delta_+}{4},\,1,\, 1- \frac{z^4}{z_H^4}\right),
\end{equation}
where $\phi_H$ is related to $c$ as
%c\,z^{\Delta_-}\; {}_2F_1\left(\frac{\Delta_-}{4},\,\frac{\Delta_-}{4},\,\frac{\Delta_-}{2},\, z^4/z_H^4\right) %+ d \, z^{\Delta_+} \;{}_2F_1\left(\frac{\Delta_+}{4},\,\frac{\Delta_+}{4},\,\frac{\Delta_+}{2},\,  %z^4/z_H^4\right),
%\end{equation}
%with
%\begin{equation} \label{eq:dc}
%d = - c\; 2^{2-\Delta_-}\,z_H^{\Delta_--\Delta_+}\, D(\Delta_-)\,,D(\Delta_-)=\frac{\pi %2^{\Delta_-}}{2-\Delta_-}\cot\left(\pi \Delta_-/4\right)\frac{\Gamma(\Delta_-/2)^2}{\Gamma(\Delta_-/4)^4}.
%\end{equation}
%The value of the dilaton at the horizon is then given by
\begin{equation}\label{eq:phi-h}
\phi_H = c \;z_H^{\Delta_-}\;\frac{\Delta_+-2}{2}\,\frac{\Gamma(\frac{\Delta_+}{4})^2}{\Gamma(\frac{\Delta_+}{2})}.
\end{equation}
The second order corrections to $B$ and $f$ can in turn be obtained by substituting $\phi(z)$ into the corresponding equations~\cite{Hohler:2009tv,Cherman:2009tw}.

\subsection{Asymptotic expansions of thermal quantities and transport coefficients}
With the asymptotic solutions, one can calculate all the thermodynamic quantities. A novel method to compute the energy density and pressure  is proposed in \cite{Hohler:2009tv}, through the derivative of the action to the boundary value of the metric.  From the derivation one finds that the parameter $f_0$ is proportional to the enthalpy $\omega=\epsilon+p$. Here, we redo the calculation starting from the black hole entropy, and then obtain the other quantities through thermodynamic relations. The results, up to quadratic order in $\phi_H$, are:
\begin{eqnarray}
s&=&\frac{2\pi l^3}{\kappa^2}\frac{1}{z_H^3},\nonumber\\
\epsilon&=&\frac{3 \omega+\theta}{4},~~~~p=\frac{\omega-\theta}{4},\nonumber\\
\omega&=&\frac{l^3}{2\kappa^2} f_0=\frac{2l^3}{\kappa^2}\frac{1}{z_H^4}\left[1+A(\Delta_+)\phi_H^2+{\cal O}(\phi_H^4)\right],\nonumber\\
\theta&=&\frac{2l^3}{\kappa^2}\frac{1}{z_H^4}\left[B(\Delta_+)\phi_H^2+{\cal O}(\phi_H^4)\right],\nonumber\\
T&=&\frac{1}{\pi z_H}\left[1+A(\Delta_+)\phi_H^2+{\cal O}(\phi_H^4)\right],
\end{eqnarray}
with
\begin{eqnarray}
A(\Delta_+)&=&-\frac{\Delta_+-2}{6\pi}\tan \left(\frac{\pi\Delta_+}{4}\right),\nonumber\\
B(\Delta_+)&=&-\frac{4-\Delta_+}{2\pi}\tan \left(\frac{\pi\Delta_+}{4}\right).
\end{eqnarray}
One can further calculate those transports coefficients in the system. The speed of sound is easily obtained from $c_s^2=\mathd p/\mathd \epsilon$, or equivalently
\begin{equation}
c_s^2=\frac{\mathd \log T}{\mathd \log s}.
\end{equation}
The ratio of the shear viscosity to the entropy density takes the universal value $1/4\pi$~\cite{Policastro:2001yc}, since Einstein gravity action is considered. The bulk viscosity has also been calculated in \cite{Cherman:2009kf} through the corresponding Green function \cite{Gubser:2008sz}. The result is confirmed in \cite{Yarom:2009mw} and generalized to arbitrary spacetime dimensions. Here we use instead the novel formula derived from the null horizon focusing equation~\cite{Eling:2011ms}, and find exactly the same result. Those coefficients are given by the asymptotic expressions
\begin{eqnarray}
c_s^2&=&\frac{1}{3}-C(\Delta_+)\phi_H^2+{\cal O}(\phi_H^4),\nonumber\\
\eta&=& \frac{l^3}{2\kappa^2}\frac{1}{z_H^3},\nonumber\\
\zeta&=&\frac{2\pi l^3}{\kappa^2}\frac{1}{z_H^3}\left[\frac{(4-\Delta_+)^2}{36\pi}\phi_H^2+{\cal O}(\phi_H^4)\right],
\end{eqnarray}
with
\begin{equation}
C(\Delta_+)=-\frac{1}{9\pi}(4-\Delta_+)(\Delta_+-2)\tan \left(\frac{\pi\Delta_+}{4}\right). \nonumber
\end{equation}
All the above expressions are obtained assuming $2<\Delta_+<4$. As shown in \cite{Hohler:2009tv,Cherman:2009tw}, under such condition $c_s^2$ is always smaller than the conformal value. A related fact is that the trace anomaly $\theta$ is always positive.  When $\Delta_+=2$, all the expressions except $\theta$ remain valid. %Since now $\phi_H\sim z_H^2\sim T^{-2}$, the enthalpy $\omega$ receives a constant correction, as in the bag model.
An additional logarithmic term appears in $\theta$:
\begin{equation}
\theta=\frac{2l^3}{\kappa^2}\frac{1}{z_H^4}\left(-\frac{8}{\pi^2}\phi_H^2\log z_H+\frac{2}{3\pi^2}\phi_H^2+{\cal O}(\phi_H^4)\right).
\end{equation}
The results for $\Delta_+=4$ can be obtained by analytically continuation. In this case, $\phi$ is zero all the way to the horizon, and no corrections appear at all. Therefore, if we insist that the dilaton is dual to the dimension-$4$ operator $\Tr F_{\mu\nu}^2$, no power corrections will be induced. The resulting thermal quantities increase quickly to the free interacting limit after the phase transition, see e.g. model III of \cite{Gubser:2008yx}\footnote{Model I and II in \cite{Gubser:2008yx} do not confine at low temperature, and show different temperature behavior from model III.} and \cite{Li:2013oda,Li2014}. Accordingly, the bulk viscosity decreases sharply above the phase transition~\cite{Gubser:2008yx}, in contrast to the observation in \cite{Meyer:2007dy,Rajagopal:2009yw}

%a little different from the bag model expectation.

As observed in~\cite{Andreev:2007zv,Andreev:2009zk,Zuo:2014vga}, such asymptotic expansions can be extrapolated to describe the theory in the intermediate temperature region. In the next section we will give two simple examples to show that this extrapolation indeed works well. In order to mimic the quadratic power behavior for the trace anomaly $\theta$ observed in lattice simulation~\cite{Pisarski:2006yk}, one has to choose $\Delta_+=3$. With such a choice, one finds $\phi_H\sim z_H\sim T^{-1}$, and thus
\begin{equation}
\theta\sim T^2.
\end{equation}
%As a manifestation of this choice, in \cite{Noronha:2009ud} it is shown that with $\Delta_+\sim 3.4$ the model quite accurately reproduces the lattice data for $\theta$.
A similar deduction can be done in 3D, with the help of the asymptotic solutions for arbitrary spacetime dimension~\cite{Yarom:2009mw}. In this case, $\Delta_+=5/2$ has to be imposed in order to generate the $1/T$ correction for the trace anomaly in 3D~\cite{Caselle:2011mn}.
%with $\Delta_+$ the dimension of the operator dual to $\phi$ and $\Delta_-=4-\Delta_+$.  According to the %dictionary, the dilaton behaves near the boundary as $\phi\sim z^{\Delta_-}$. Moreover, at high temperature, %the temperature is related to the horizon as
%\begin{equation}
%T\sim\frac{1}{\pi z_H}.
%\end{equation}

\subsection{Asymptotic expansion of the quark free energy}
Now, let us turn to the heavy quark free energy and the Polyakov loop. The derivative of the quark free energy, with respect to the temperature, has a simple expression in this system. It is related to the speed of the sound  and to the potential of the dilaton as~\cite{Noronha:2009ud}
\begin{equation}
\frac{\mathd F_Q}{\mathd T}=4\pi T_s \frac{\mathe ^{\sqrt{2/3}\phi_H}}{V(\phi_H)}\frac{1}{c^2_s(\phi_H)}.\label{eq.dFdT}
\end{equation}
 Thus, this quantity would diverge where the velocity of sound vanishes. This  happens at the minimum temperature $T=T_{\rm{min}}$, if the zero temperature background is confining~\cite{Gursoy:2008za,Gursoy:2008bu}.
Since the deconfining temperature is always above $T_{\rm{min}}$, $c^2_s$ is always positive in the deconfining phase and $\frac{\mathd F_Q}{\mathd T}$ would never diverge.
Actually, the deconfining temperature turns out to be very close to $T_{\rm{min}}$, and the resulting $c^2_s$ is quite small at the critical point~\cite{Gubser:2008yx}. Lattice simulations  show this kind of behavior of the square of the velocity of sound at $T_c$ \cite{Boyd:1996bx}.

As shown before, the potential of the dilaton has the expansion near the boundary, or for small $\phi$
\begin{equation}
V(\phi)\sim -\frac{12}{l^2}+\frac{1}{2l^2}\Delta_+(\Delta_+-4)\phi^2 +{\cal O}(\phi^4).\label{eq.Vexp}
\end{equation}
Taking all the factors into account, one finds:
\begin{equation}
\frac{\mathd F_Q}{\mathd T}=-\frac{l^2}{2l_s^2}\left[1+\sqrt{\frac{2}{3}}\phi_H+\left(\frac{1}{3}+\frac{1}{24}\Delta_+(\Delta_+-4)+3C(\Delta_+)\right)\phi_H^2+{\cal O}(\phi_H^3)\right]. \label{eq.dFdT1}
\end{equation}
Again, we could extrapolate such an expansion to the intermediate temperature region, as done in ~\cite{Andreev:2007zv,Andreev:2009zk,Zuo:2014vga}. Choosing $\Delta_+=3$, the leading correction to $\mathd F_Q/\mathd T$, and also the renormalized Polyakov loop, are linear in the inverse temperature. As a result, the renormalized Polyakov loop approaches the asymptotic value much more slowly, which is indeed observed in~\cite{Noronha:2009ud}. In order to generate the quadratic contribution observed for the Polyakov loop, $\Delta_+=2$ has to be imposed, in contrast to the previous choice. Indeed, with such a choice the lattice data for the Polyakov loop can be well described~\cite{Li:2011hp}. However, the thermodynamic quantities, in particular the trace anomaly, are not well reproduced at the same time~\cite{Li:2011hp}. Therefore, in the present framework it seems impossible to accommodate the lattice data of both the trace anomaly and the renormalized Polyakov loop. The reason can be attributed to the fact that one has to use the string frame metric to calculate the quark free energy and the Polyakov loop, which thus depend linearly on the dilaton. While in calculating the thermodynamic quantities, the Einstein frame metric is used, which achieves back reaction from the dilaton only at even orders.

\section{Two simple models}

\subsection{Model I: $\Delta_+=3$}
\subsubsection{Background functions}
In this Section we choose a special solution of the Einstein-dilaton system to confirm the general behavior discussed above. In principle, one should start with a choice of the dilaton potential and then solve the coupled equations of the system. For simplicity, we choose to start from a special form of the metric, instead. Through the equation of motion, this special metric will correspond to a special dilaton potential, which may not be of a simple form. The potential obtained in this way will depend on the black hole horizon, but the dependence is found to be very weak. Instead of (\ref{eq:ansatz1}),  we choose to work with the slightly different ansatz~\cite{Gursoy:2008za},
\begin{equation}
ds^2=b^2(r)\left[-f(r)dt^2+d{\bf x}^2+{dr^2\over f(r)}\right],
\label{eq:ansatz2}
\end{equation}
for which the equations of motion read:
\begin{eqnarray}
&& 6{\dot b^2\over b^2}-3{\ddot b\over b}={\frac{4}{3}} \dot\Phi^2,\label{eq.eq2}\\
&&{\ddot f\over \dot f}+3{\dot b\over b}=0,\label{eq.eq3}\\
&&6{\dot b^2\over b^2}+3{\ddot b\over b}+3{\dot b\over b}{\dot f\over f}=-{b^2\over f}V(\Phi),\label{eq.eq1}
\end{eqnarray}
where $\dot y\equiv \mathd y/\mathd r$, and $\Phi=\sqrt{3/8}~\phi$.
The solutions to  these equations have been systematically studied for different dilaton potentials~\cite{Gursoy:2007cb,Gursoy:2007er}.  An important observation is that linear confinement, as proposed in the soft-wall model \cite{Karch:2006pv,Andreev:2006vy}, can only be obtained when the potential has the infrared behavior $V(\Phi)\sim \Phi^{1/2} \exp [\frac{4\Phi}{3}]$. The metric function $b(r)$ will then be of $\exp{(-r^2)}$ in the infrared region. Combining this with the asymptotic Anti-de Sitter metric in the ultraviolet, one can choose to work with the following simple ansatz for $b(r)$:
\begin{equation}
b(r)=\frac{l}{r} \exp \left(-\frac{r^2}{R^2}\right).\label{eq.metric}
\end{equation}
%The function in the exponential has been chosen to be quadratic in $r$, in order to reproduce expected Regge %trajectories in the model~\cite{Gursoy:2007er}.
%Now we will show that this choice also gives rise to the quadratic power corrections for the thermodynamic quantities.
It is easy to see that the infrared deformation induces quadratic corrections in the ultraviolet, in a similar manner as the fuzzy bag. Since the ultraviolet corrections of the metric are square of that of the dilaton, the leading UV behavior of the dilaton should be $r$ and the dual dimension  $\Delta_+=3$.
The zero temperature solution with this ansatz has been given in ref.~\cite{Gursoy:2007er}. Recently, this kind of ansatz, with varying power index in the exponential, has also been employed to
study the thermodynamics in 3D~\cite{Caselle:2011mn}. %As commented in ~\cite{Caselle:2011mn}, the solution with such a %metric will not be exact since the dilaton potential will depend on the black-hole horizon when $\phi$ is  large. %However, the qualitative picture obtained from such a solution will still be reasonable.
With this choice, the dilaton and the black-hole factor are:
\begin{eqnarray}
\Phi(r)&=&\frac{3}{2}\frac{r}{R}\sqrt{\frac{3}{2}+\frac{r^2}{R^2}}+\frac{9}{4}\log\left[\sqrt{\frac{2}{3}}\frac{r}{R}+\sqrt{\frac{2r^2}{3R^2}+1}\right],\nonumber\\
f(r)&=&1-\frac{1+\mathe^{3r^2/R^2}(3r^2/R^2-1)}{1+\mathe^{3r_H^2/R^2}(3r_H^2/R^2-1)}.
\end{eqnarray}
All these functions then completely fix the background, which we call model I.
The so-called thermal superpotential $W$, defined as $W\equiv -9 \dot b/(4 b^2)$, is given by
\begin{equation}
W(r)=\frac{9}{4l}(1+\frac{2r^2}{R^2})\exp\left[\frac{r^2}{R^2}\right].
\end{equation}
The dilaton potential can also be obtained, though the expression is quite involved. The infrared expansion of the potential in term of $\Phi$ is:
\begin{equation}
V(\Phi)\sim -\frac{27}{4l^2} \mathe^{-3/2} \Phi^{1/2} \exp \left[\frac{4\Phi}{3}\right], \quad \Phi \to \infty.
\end{equation}
As shown in Refs.~\cite{Gursoy:2007cb,Gursoy:2007er,Gursoy:2008za}, such a large-$\phi$ behavior of the potential is required to reproduce the infrared form of the metric in Eq.(\ref{eq.metric}).
%Thermodynamics in the Einstein-dilaton system with such a dilaton potential have also been studied in %\cite{Gubser:2008yx,Gubser:2008ny}.
In the ultraviolet, the potential can be expanded as
\begin{equation}
V(\Phi)\sim -\frac{12}{l^2}-\frac{4}{l^2} \Phi^2, \quad  \Phi \to 0.
\end{equation}
Comparing with Eq.~(\ref{eq.Vexp}), one finds that the dimension of the operator dual to $\Phi$ is $\Delta_+=3$. One can also check this by noting that $\Phi\sim r$ near the boundary. An artifact of our choice of the metric is that the coefficient of the leading infrared term of the potential is completely constrained by the ultraviolet properties, as in \cite{Gursoy:2007er,Gursoy:2008bu}.

\subsubsection{Phase transition and  critical temperature}
With all these functions we can calculate the thermodynamic quantities. The temperature is determined by
\begin{equation}
4\pi T=-f'(r_H),
\end{equation}
and the explicit form is
\begin{equation}
T=\frac{9 ~r_H^3}{2\pi R^4} \left[\mathe^{-3 r_H^2/R^2}+3r_H^2/R^2-1\right]^{-1}.
\end{equation}
As shown in ref.~\cite{Gursoy:2008za}, one finds  a minimum temperature, $T_{\rm{min}}\simeq 0.686 R^{-1}$, with the corresponding horizon $r_{\rm{min}}\simeq 0.85 R$. Above $T_{\rm{min}}$ there are two different horizon values, corresponding to two different kind of black hole solutions. When the horizon is close to the boundary, $0<r_H<r_{\rm{min}}$, we have a big black-hole, and the temperature scales with the horizon as
\begin{equation}
T\sim \frac{1}{\pi r_H}[1+\frac{r_H^2}{R^2}+{\cal O}(r_H^4)].
\end{equation}
When the horizon is in the deep infrared, we get a small black-hole, with  temperature
\begin{equation}
T\sim \frac{3~r_H}{2\pi~R^2}+{\cal O}(r_H^{-1}).
\end{equation}
The entropy can be read for the metric on the horizon
\begin{equation}
s=\frac{l^3}{4G_5}\frac{1}{r_H^3} e^{-3r_H^2/R^2}.\label{eq.entropy}
\end{equation}
Integrating the entropy over $T$, one gets the pressure (after subtracting that of the thermal AdS):
\begin{equation}
p=\int^T s(T) \mathd T = -\frac{l^3}{4G_5} \int_{r_H}^\infty  \frac{1}{r_H^3} \exp\left[-3r_H^2/R^2\right] \frac{\mathd T}{\mathd r_H} \mathd r_H.\label{eq.pressure}
\end{equation}
As analyzed in detail in Ref.~\cite{Gursoy:2008za}, the above expression is valid for both the big and small black-hole. Notice that the integration constant in the pressure is fixed by the requirement that the small black-hole asymptotically reproduces  the thermal gas solution in the high-$T$ limit. When taking the limit $T\to \infty$ for the big black-hole, one recovers the asymptotic behavior
\begin{equation}
p(T)\to \frac{\pi^3l^3}{16G_5} T^4+{\cal O}(T^2).
\end{equation}
Matching it to the free field limit result, one can fix the constant:
\begin{equation}
\frac{l^3}{4G_5}=\frac{4}{45\pi} (N_C^2-1).
\end{equation}

For a fixed temperature, the pressure of the big black-hole is always bigger, signaling that it is always preferred than the small black-hole. The transition between the big black-hole and the thermal gas solution occurs when the pressure vanishes, which describes the deconfinement phase transition~\cite{Witten:1998zw}. The critical temperature is $T_c= 0.693~ R^{-1}$, slightly above the minimum temperature. A rough estimate of the critical temperature can be derived form the property of the string frame warp factor $b_s=b~\mathe^{2\Phi/3}$~\cite{Kinar:1998vq}. In a confining background, $b_s$ should develop a nonzero minimum at some point $r_*$. The gravitational repulsion from the infrared region beyond this point confines the string attached at the boundary. The phase transition occurs when the black-hole horizon coincides with this point~\cite{Andreev:2006eh}. In the present model, this point is at $r_*=R/\sqrt{2}$ and the corresponding critical temperature is $T_*\simeq 0.707 R^{-1}$. One finds the three temperatures are very close to each other, with  $T_{\rm{min}}\lesssim T_c\lesssim T_*$.

It would be interesting to make a numerical evaluation of the critical temperature. As investigated in \cite{Gursoy:2008za}, in the large $\Phi$ region, the thermal solution for the scale factor $b(r)$ and the dilaton are very well approximated by the zero temperature form. From the infrared properties of the background of the latter, one can then deduce the asymptotic form of the Schr\"{o}dinger potential for different spin modes. It turns out that the potential for the scalar glueballs, tensor glueballs and also the vector mesons, the Schr\"{o}dinger potential takes the same asymptotic form:
\begin{equation}
V_S(r)\sim \frac{9}{4} \frac{r^2}{R^4}+{\cal O}(1).
\end{equation}
The resulting spectrum is then
\begin{equation}
m_n^2\sim 6 n R^{-2}.
\end{equation}
With the experimental fit for the vector meson spectrum $m_n^2\sim 0.93~n ~\mbox{GeV}^2$~\cite{Eidelman:2004wy}, one fixes $R=2.54~\mbox{GeV}^{-1}$,  obtaining the critical temperature
\begin{equation}
T_c\simeq 273 ~\mbox{MeV}.
\end{equation}
%Later we will always use this value unless specified especially.
Interestingly, this value is very close to the large-$N$ lattice prediction for pure SU(N) gauge theory~\cite{Lucini:2005vg}~\footnote{In our model no dynamical quarks are introduced.}.
One can also evaluate the confining string tension in the quark potential with the zero-temperature background. Following \cite{Kinar:1998vq,Andreev:2006ct,Gursoy:2007er}, this is given by
\begin{equation}
\sigma=\frac{1}{2\pi l_s^2}~b_s(r_*)^2\simeq 4.5 ~\frac{l^2}{l_s^2} ~\frac{1}{R^2}.
\end{equation}
In order to produce the lattice result $\sigma\simeq (440\mbox{MeV})^2$, one finds
\begin{equation}
\frac{l^2}{l_s^2}\simeq 1/3.6~.\label{eq.ls}
\end{equation}
Thus, the string length is of the same order as the asymptotic AdS radius, indicating that it is an effective quantity in 5 dimension. Similar results are also obtained in \cite{Andreev:2006ct,Gursoy:2009jd}. Since our later calculation of the Polyakov loop is very similar to the quark potential, we will use this relation as an  input. It is worth noticing that a very close value for this ratio is used in ~\cite{Noronha:2009ud} to fit the lattice data for the renormalized Polyakov loop.

\subsubsection{Trace anomaly}

%Lattice study shows that in 3 dimension the trace anomaly is also dominated by a $T^2$ term~\cite{Caselle:2011mn}, which %means that a similar expansion as (\ref{interpolating_formula_for_Delta1}) is dominated by a linear term in inverse temperature.

From the expressions (\ref{eq.entropy},\ref{eq.pressure}) for $s$ and $p$, it is  straightforward to obtain the trace anomaly $\theta=\epsilon-3p=Ts-4p$. As expected, the anomaly can be expressed as a series of the inverse temperature squared, with a leading $T^2$ term
\begin{equation}
\theta = \frac{\pi^2}{45} (N_C^2-1) \cdot  \left( 5.06 \, T_c^2 \, T^2 +{\cal O} \left(\log\left(\frac{T}{T_c}\right)\right)\right).\label{eq.theta3}
\end{equation}

\begin{figure}[ht]
\centering
	\includegraphics[width=0.7\textwidth]{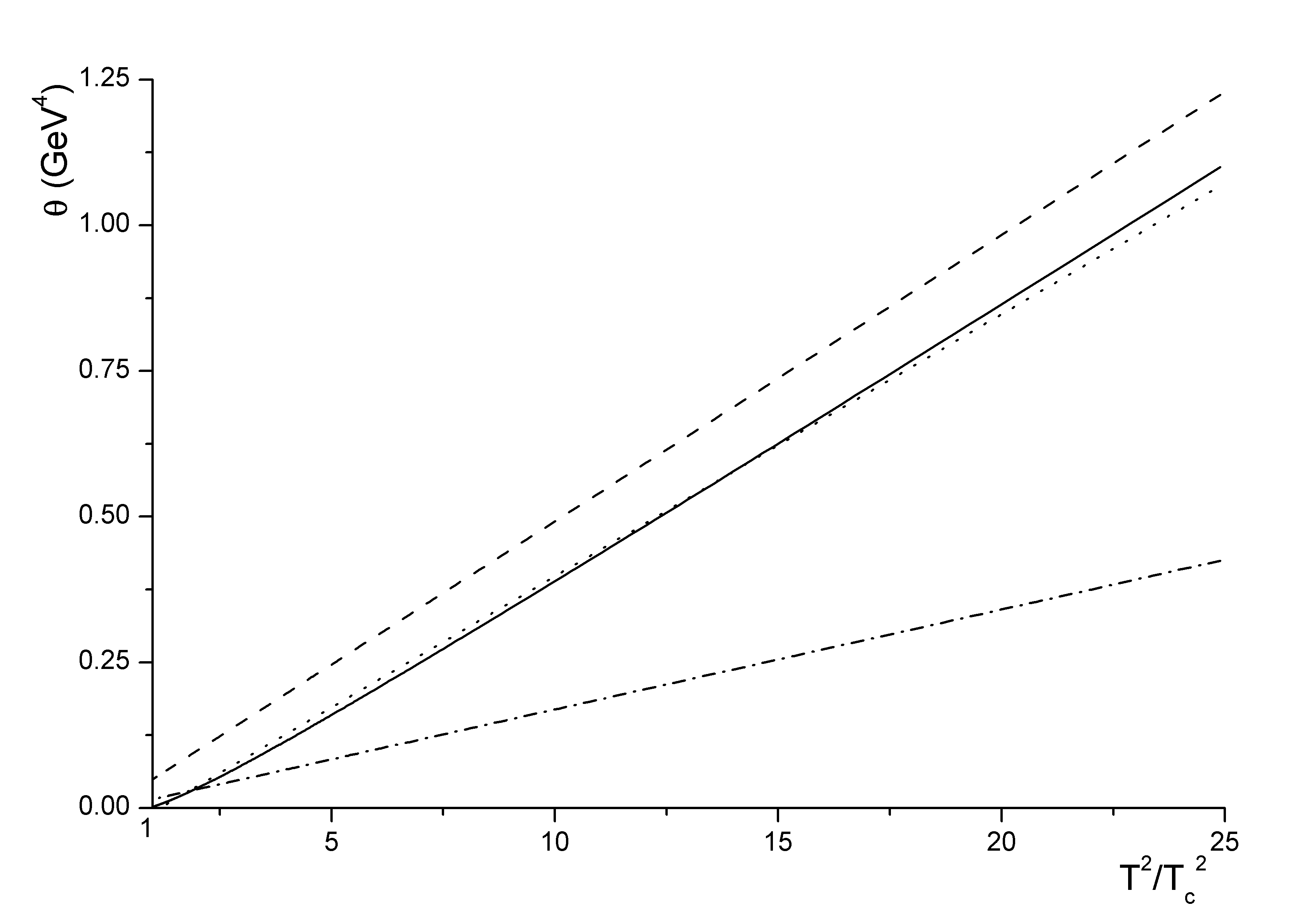}
%\vspace*{8pt}
\caption{\it Holographic results for the trace anomaly versus $T^2/T_c^2$. The exact results are given by the solid line, while the leading term results are the dashed line and the linear fit is the dotted line. Also plotted is the lattice fit (\ref{eq.theta0}) from \cite{Panero:2009tv}~(dash-dotted line).}\label{fig:Anomaly}
\end{figure}

In FIG.~\ref{fig:Anomaly} we plot the exact results in the range of temperature  $T_c<T<5~T_c$~(solid line), together with the approximated ones with only the $T^2$ term~(dashed line). Except the region close to $T_c$, the anomaly is indeed dominated by the $T^2$ term, and thus it shows a linear pattern as the lattice result (\ref{eq.theta0})~(dash-dotted line). Indeed, the result can be quite well fitted by the form (\ref{eq.theta1}) (dotted line), with the fitted parameter
\begin{equation}
f_3=4.61,~~~~f_4=-5.39.
\end{equation}
Similar to  lattice data, $f_4$ turns out to negative, making the actual results smaller than the leading term. However, the absolute values of both parameters are much bigger than those from lattice simulation, and the constant term turns out to be of the same order as the leading term. This is not  surprising, since in our simple ansatz (\ref{eq.metric}) the UV power corrections are completely inherited from the
infrared exponential form. By fine-tuning the metric in the ultraviolet term by term, one could reproduce the lattice data quantitatively. Alternatively, one could make a tuned choice of the dilaton potential as in \cite{Gursoy:2009jd} to reproduce the lattice data for thermal quantities.

\subsubsection{Polyakov loop and quark free energy}
Let us check if such a simple form can be reproduced in the present model, in the same way as for the trace anomaly. To do this, the expression (\ref{eq.dFdT}) cannot be directly used since we do not have the explicit form the dilaton potential. Following the derivation in \cite{Andreev:2009zk}, one finds
\begin{equation}
F_Q(T)=\frac{l^2}{2\pi l_s^2}\int_0^{r_H}\frac{1}{r^2}\exp \left[-\frac{2r^2}{R^2}+\frac{4}{3}\Phi(r)\right] \mathd r.
\end{equation}
As observed in \cite{Noronha:2009ud}, such an expression gives $F_Q(T)\sim -\frac{l^2}{2\pi l_s^2}~ T$ at large $T$, due to the asymptotic AdS behavior. Since the renormalized Polyakov loop should approach one  in the large-$T$ limit, the renormalized Polyakov loop could therefore be defined by subtracting the result in thermal AdS:
\begin{equation}
F^R_Q(T)=\frac{l^2}{2\pi l_s^2}\int_0^{r_H}\frac{1}{r^2}\left[\exp \left(-\frac{2r^2}{R^2}+\frac{4}{3}\Phi(r)\right)-1\right] \mathd r.\label{eq.F_R}
\end{equation}
However, in the present case with $\Phi(r)\sim r$ in the UV, such a subtraction is not enough to eliminate the whole divergence, indicating some physical inconsistence. Due to this, we choose to study the derivative of the quark free energy
\begin{equation}
\frac{\mathd F^R_Q(T)}{\mathd T}=\frac{l^2}{2\pi l_s^2}\frac{1}{r_H^2}\left[\exp \left(-\frac{2r_H^2}{R^2}+\frac{4}{3}\Phi(r_H)\right)-1\right]\left[\frac{\mathd T}{\mathd r_H}\right]^{-1},\label{eq.dF_R}
\end{equation}
which vanishes in the high temperature limit. This can  be evaluated with the numerical value for the ratio $l^2/l_s^2$ (\ref{eq.ls}). If a $T_c^2/T^2$ term exists and makes the dominant contribution in the medium temperature region, the combination $(T^2/T_c^2)~\mathd F^R_Q(T)/\mathd T$ should approach a constant.

% in comparison with the lattice fit for SU(4) and SU(5) gauge theory.
\begin{figure}[ht]
\centering
	\includegraphics[width=0.8\textwidth]{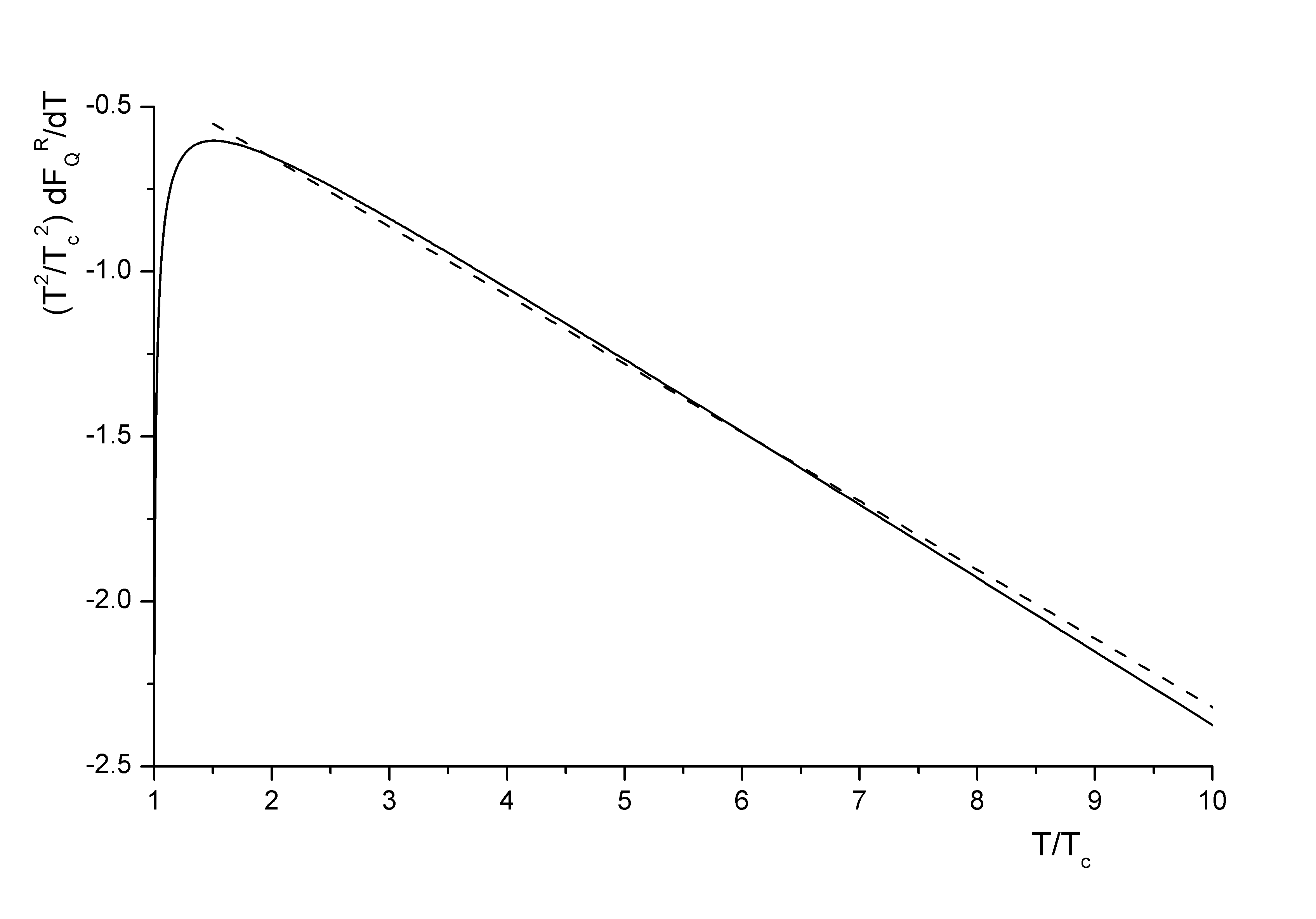}
%\vspace*{8pt}
\caption{\it Holographic results for combination of the derivative of the quark free energy with $T^2/T_c^2$ in model I. The linear fit in the region $1.5~T_c<T<10~T_c$ is shown in dashed line.}\label{fig:dFdT}
\end{figure}

In FIG. \ref{fig:dFdT} we show the results of this combination versus $T/T_c$. Instead of a plateau, we find a linear decreasing of this combination, which can be  fitted in the region $1.5~T_c<T<10~T_c$ by
\begin{equation}
(T^2/T_c^2)~\frac{\mathd F^R_Q(T)}{\mathd T}\approx -0.21~\frac{T}{T_c}-0.24.
\end{equation}
As we expected from the analysis in the previous section, the result is dominated by a linear term, though the quadratic contribution is nonvanishing. This is in accordance with our  expansion (\ref{eq.dFdT1}). Therefore, we confirms that, in the Einstein-dilaton system, we cannot reproduce the lattice data for the quark free energy, and neither the renormalized Polyakov loop. Such a conclusion is independent of the renormalization scheme of the Polyakov loop. Adding a constant to the quark free energy could induce a linear term in the renormalized Polyakov loop, but does not affect the derivative of the free energy.

%The result can be fitted quite well with a quadratic polynomial
%\begin{equation}
%\frac{\mathd F_Q(T)}{\mathd T}\simeq \frac{l^2}{2\pi \alpha'} (a_0+a_1\frac{T_c}{T}+a_2\frac{T_c^2}{T^2}).
%\end{equation}
%The fitted parameters, for the temperature region $3~T_c<T<14~T_c$, are
%\begin{equation}
%a_0=-3.21,\quad a_1=-5.92,\quad a_2=-12.9.
%\end{equation}
%These values are very close to those appearing in the asymptotic expansion
%\begin{equation}
%\frac{\mathd F_Q(T)}{\mathd T}\simeq \frac{l^2}{2\pi \alpha'} (-3.14-7.07\frac{T_c}{T}-7.29\frac{T_c^2}{T^2}+{\cal %O}(\frac{T_c^3}{T^3})).
%\end{equation}
%If one chooses to fit the curve in a lower temperature region, those parameters differ from the asymptotic ones more.
%In contrast to the lattice fit, our results shows no evidence of a linear pattern.

%\newpage
\subsection{Model II: $\Delta_+=2$}
\subsubsection{Background functions}
In the previous model the quadratic terms appear in the metric, while the dilaton is linear in UV. Now we consider another choice, with the quadratic term coming directly from the dilaton. As a result, the metric corrections will  be quartic. In the infrared we  still require $b(r)\sim\exp{(-r^2)}$ to ensure linear confinement. A simple ansatz with both properties is
\begin{equation}
b(r)=\frac{l}{r} \left(1+\frac{r^2}{R^2}\right)~\exp \left[-\frac{r^2}{R^2}\right].\label{eq.metric2}
\end{equation}
The dilaton and the black-hole factor can be solved:
\begin{eqnarray}
\Phi(r)&=&-\frac{3}{8}\left(2\sqrt{10}-2\sqrt{10+\tilde{r}^2+4\tilde{r}^4}-\mbox{ArcSinh}\left(\frac{3}{\sqrt{31}}\right)+ \mbox{ArcSinh}\left(\frac{3+4\tilde{r}^2}{\sqrt{31}}\right)\right.\\\nonumber
&&\left.+4\sqrt{2}\log\left[\frac{7-\tilde{r}^2+4\sqrt{5+\tilde{r}^2+2\tilde{r}^4}}{(7+4\sqrt{5})(1+\tilde{r}^2)}\right]\right)\\
f(r)&=& 1-\frac{A(\tilde{r})-A(0)}{A(\tilde{r}_h)-A(0)},
\end{eqnarray}
where $\tilde{r}=r/R$, ArcSinh$(x)$ is the inverse hyperbolic Sine function and $A(\tilde{r})$ is defined through the exponential integral function Ei$(x)$ as
\begin{equation}
A(\tilde{r})=\frac{\mathe ^{3(1+\tilde{r}^2)}(2+\tilde{r}^2)-3(1+\tilde{r}^2)^2 \mbox{Ei}[3(1+\tilde{r}^2)]}{4\mathe^3 (1+\tilde{r}^2)^2}.
\end{equation}
Expanding the dilaton around $r\sim 0$ gives
\begin{equation}
\Phi(r)\sim \frac{3}{2}\sqrt{\frac{5}{2}}~\tilde{r}^2-\frac{21}{8\sqrt{10}}~\tilde{r}^4+{\mathcal O}(r^6).
\end{equation}
Therefore, the dual dimension is indeed $\Delta_+=2$. This can be also be confirmed by solving the potential in the ultraviolet
\begin{equation}
V(\Phi)\sim -\frac{12}{l^2}-\frac{16}{3l^2} \Phi^2, \quad  \Phi \to 0.
\end{equation}
The infrared form of the super potential and dilaton potential are the same as model I, since the leading exponential behavior in $b(r)$ remains unchanged. Correspondingly, the asymptotic hadron spectrum is the same as in model I,  together with the value  of the parameter $R$. A similar model with $\Delta_+=2$ has been studied in \cite{Li:2011hp}, though the infrared background is quite different. Later, we shall see that the results for the thermodynamic quantities and the Polyakov loop show a similar pattern as ours.

\subsubsection{Phase transition and the critical temperature}

The temperature can be expressed through the function $A(\tilde{r})$ defined before
\begin{equation}
T=\frac{1}{4\pi R}\frac{\tilde{r}_H^3\mathe ^{3\tilde{r}_H^2}}{(1+\tilde{r}_H^2)^3}\frac{1}{A(\tilde{r}_H)-A(0)}, \label{eq.temperature2}
\end{equation}
while the infrared dependence on $r_H$ is still
\begin{equation}
T\sim \frac{3~r_H}{2\pi~R^2}+{\cal O}(r_H^{-1}),
\end{equation}
the temperature now achieves only quartic corrections in the ultraviolet
\begin{equation}
T\sim \frac{1}{\pi r_H}[1+{\mathcal O}(r_H^4)].
\end{equation}
The minimum temperature appears now at $r_{\rm{min}}\sim 0.96~R$, with $T_{\rm{min}}\sim 0.47~R^{-1}$.
A detailed calculation of the pressure shows that the phase transition occurs when $r_c\sim 0.86~R$, with $T_c\sim0.48~R^{-1}\approx 189~\mbox{MeV}$. Such a value is very close to the lattice result of QCD~\cite{Karsch:2006xs}. Comparing with model I, we see that the critical temperature is sensitive to the quadratic terms in the metric.

We may estimate of the critical temperature from the metric factor in the string frame $b_s=b~\mathe^{2\Phi/3}$. The minimum of $b_s(r)$ occurs at $r_*\sim0.77~R$, with $b_s(r_*)\sim 2.56 ~l/R$. The corresponding temperature  is $T_*\sim 0.49~R^{-1}$. One sees again that $T_{\rm{min}}\lesssim T_c\lesssim T_*$. The minimum of $b_s(r)$ can also be used to determine the ratio $l^2/l_s^2$ from the confining string tension, giving $l^2/l_s^2\sim 1.195$.

\subsubsection{Trace anomaly}
\begin{figure}[ht]
\centering
	\includegraphics[width=0.7\textwidth]{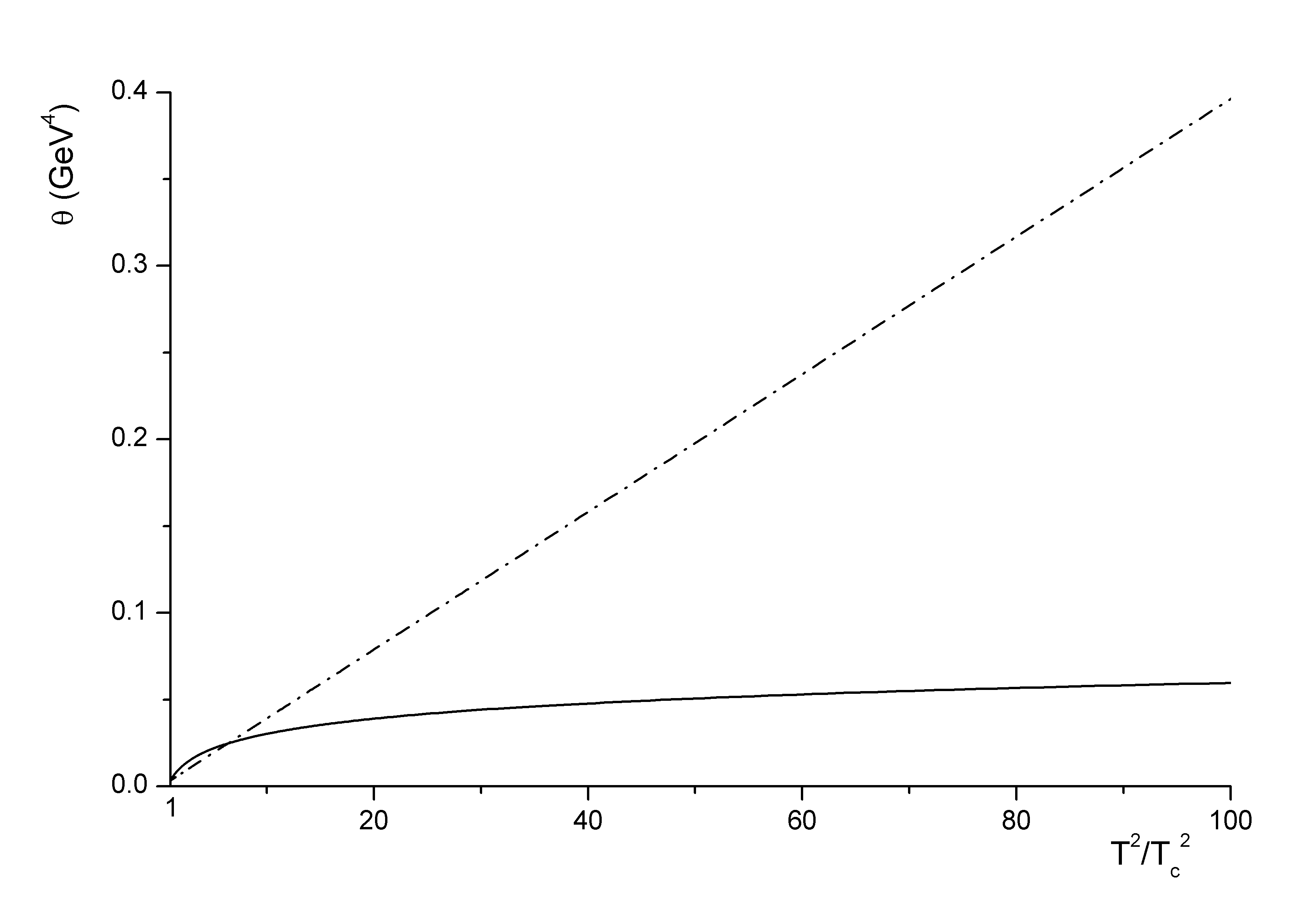}
%\vspace*{8pt}
\caption{\it Holographic results for the trace anomaly versus $T^2/T_c^2$ in model II. Also plotted is the lattice fit (\ref{eq.theta0}) from \cite{Panero:2009tv}~(dash-dotted line).}\label{fig:Anomaly2}
\end{figure}

 With the metric function (\ref{eq.metric2}) and the temperature expression (\ref{eq.temperature2}) we again calculate the entropy, pressure and  the trace anomaly. The result for the trace anomaly is plotted versus $T^2/T_c^2$ is FIG.~\ref{fig:Anomaly2}, together with the lattice fit. Compared to FIG.~\ref{fig:Anomaly}, the linear pattern of the anomaly is lost in this model, as expected. Instead, a small, almost constant contribution appears, which indicates that in the model the quartic term dominates the anomaly.

\subsubsection{Polyakov loop and quark free energy}
\begin{figure}[ht]
\centering
	\includegraphics[width=0.8\textwidth]{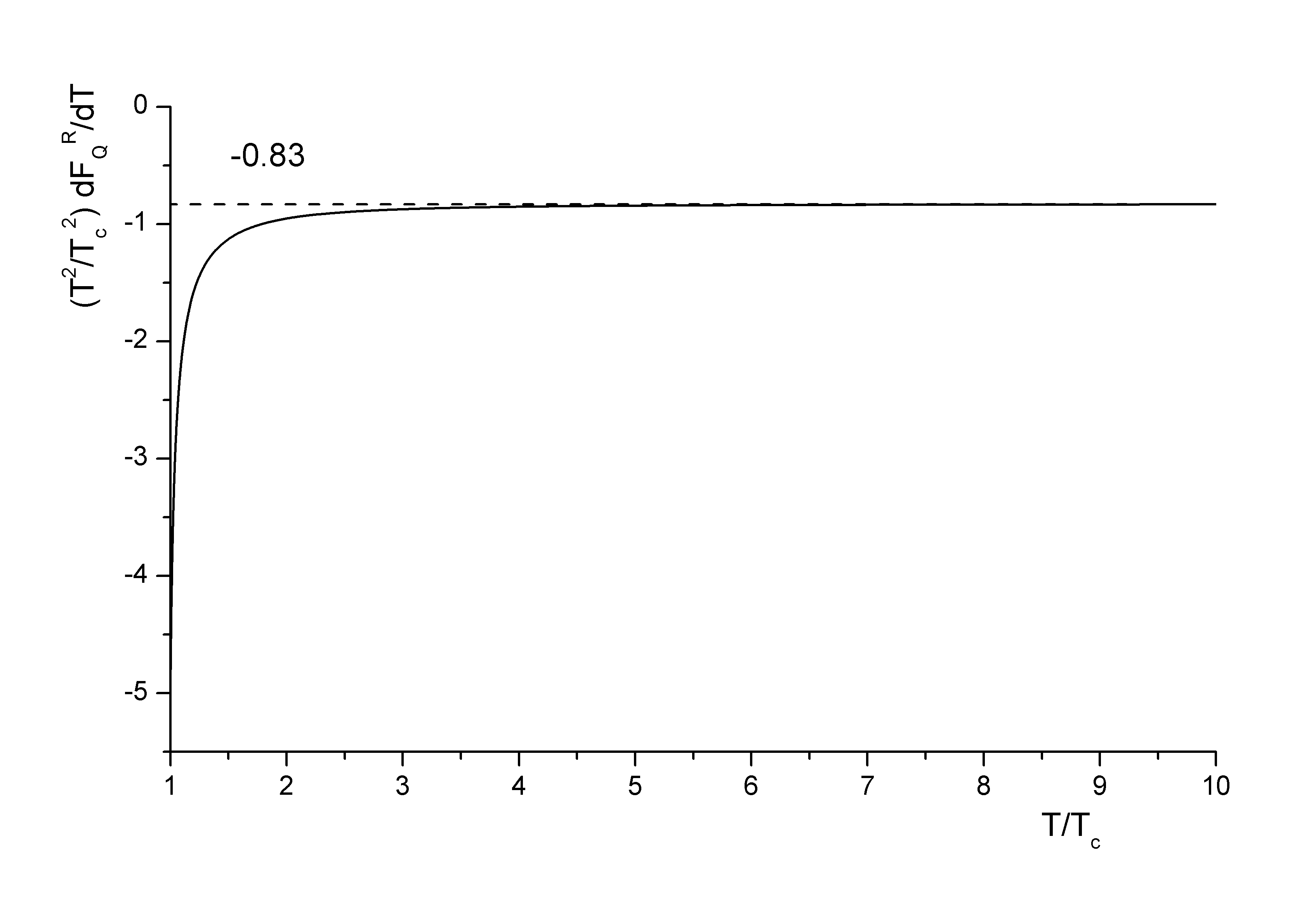}
%\vspace*{8pt}
\caption{\it Holographic results for combination of the derivative of the quark free energy with $T^2/T_c^2$ in model II. The asymptotic value is given in dashed line.}\label{fig:dFdT2}
\end{figure}

Now we consider the Polyakov loop and the quark free energy. In this case the quark free energy is  less divergent. A simple subtraction as in (\ref{eq.F_R}) is enough to eliminate the divergence. To compare with the results in model I, we first calculate the derivative of the free energy with (\ref{eq.dF_R}). From the asymptotic expansion one expects that the combination  $(T^2/T_c^2)~\mathd F^R_Q(T)/\mathd T$ approaches a constant when the temperature is large enough. Indeed, we see the appearance of a plateau in FIG.~\ref{fig:dFdT2} starting from temperature as low as $1.5~T_c$. This proves that the dominance of the quadratic term is valid in a large temperature region, even close to the critical temperature. The deviation from the asymptotic value becomes obvious only in a narrow temperature region $T_c<T\lesssim 1.5~T_c$, where more and more higher power terms start to make sizeable contributions. Due to this, the magnitude of $\mathd F^R_Q(T)/\mathd T$ becomes larger and larger, and would be divergent if one extrapolates to the minimum temperature $T_{\rm{min}}<T_c$.
\begin{figure}[ht]
\centering
	\includegraphics[width=0.7\textwidth]{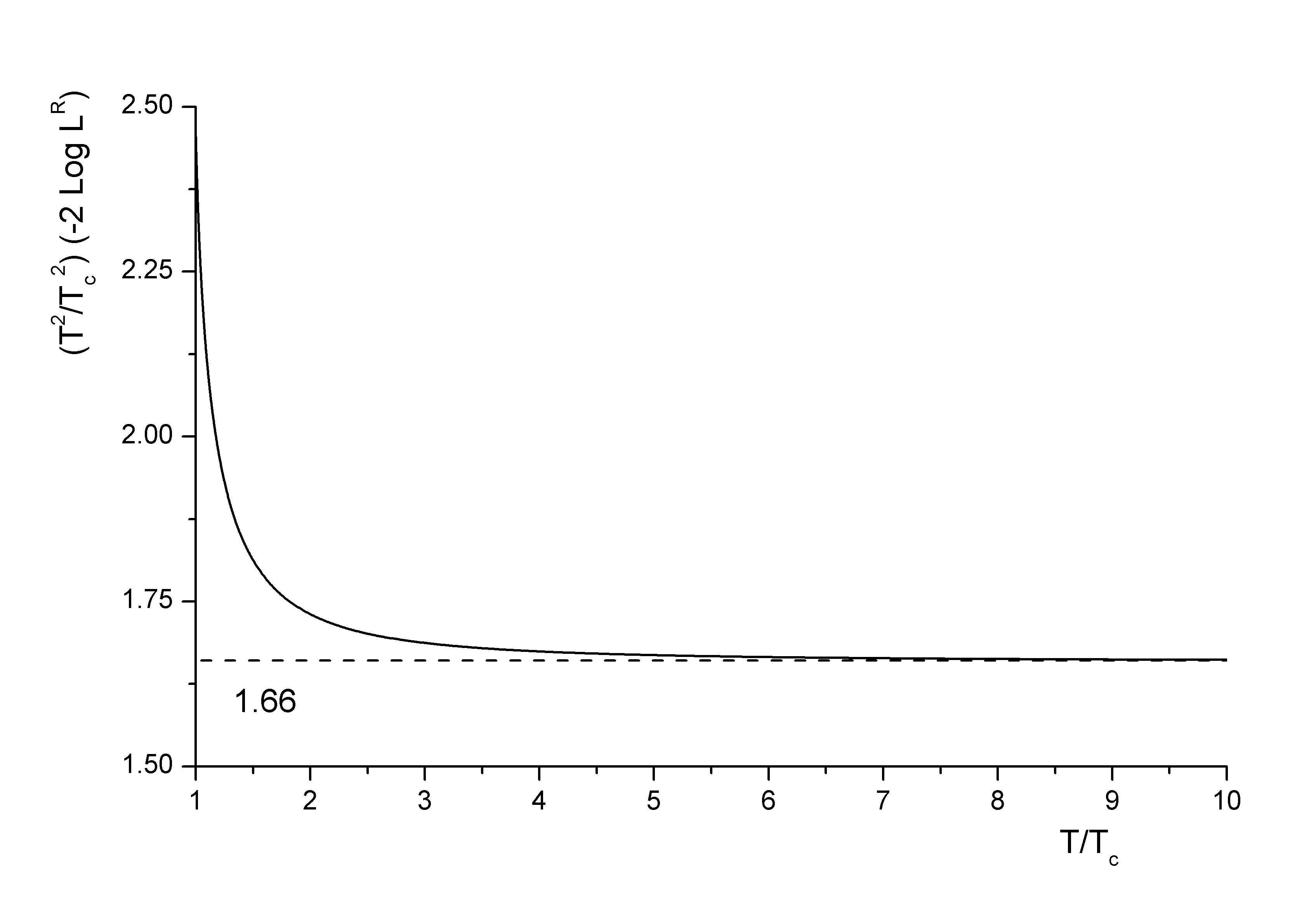}
%\vspace*{8pt}
\caption{\it Holographic results for combination of the logarithmic of the renormalized Polyakov loop, $-2 \log L^R$,  with $T^2/T_c^2$ in model II. The asymptotic value is given in dashed line.}\label{fig:Polyakov}
\end{figure}

Since now the quark free energy is well defined, we go on to study the Polyakov loop expectation value. In order to see the importance of the quadratic term, we plot the result of the combination $-2 \log L^R$ with $T^2/T_c^2$ in FIG.~\ref{fig:Polyakov}. Similar as FIG.~\ref{fig:dFdT2}, this combination approaches the asymptotic value very quickly, starting from $T\sim 2~T_c$. The asymptotic value of such a combination gives the prediction of the coefficient $b$ of the quadratic term defined in (\ref{eq.logL}), $b\sim 1.66$. Such a value is very close to the fitted values from lattice data for SU$(3)$~\cite{Megias:2005ve} and SU$(5)$~\cite{Mykkanen:2012ri} gauge theory, and little larger than the one in SU(4)~\cite{Mykkanen:2012ri}.

%\newpage
\section{Discussion}
In \cite{Zuo:2014vga} we have studied the quadratic thermal contributions with the example of ${\mathcal N}=4$ super Yang-Mills gauge theory on $S^1\times S^3$. In contrast to the case when the boundary is flat, the theory confines at low temperature and exhibits a first-order deconfinement phase transition. At the meantime, the quadratic terms appear in all the thermal quantities, and also in the Polyakov loop. It thus seems that such contributions are completely due to global change of the bulk spacetime, rather than some local fields. Such an observation is consistent with the field theory expectation, since no gauge-invariant dimension-2 operator exists.%similar as the argument for the quadratic corrections at zero temperature~\cite{Zakharov1998}.

In this work we try to generate such contributions from a local field, the dilaton. However, the thermal quantities and the Polyakov loop depend differently on the dilaton. Due to this, the quadratic terms in all of them simultaneously are not generated. Is this another indication that the quadratic thermal terms indeed reflect some global or non-local effects? Still, there could be another way out. Suppose these contributions are from the fluctuations of another filed. %If such a field couples to the metric and the dilaton in a similar way, quadratic fluctuations will in turn be generated %in both of them. These fluctuations finally appear in
Such a field will affect the thermodynamics and the Polyakov loop indirectly through the coupling to the metric and the dilaton. If the thermal quantities and the Polyakov loop depend similarly on this field, the quadratic terms in them can be induced simultaneously. To see if such a mechanism works or not, we must introduce another relevant field into the gravity-dilaton system. We plan to investigate this in the future.

\section*{Acknowledgments}

This work is partially supported by the National Natural Science Foundation of China under Grant No. 11135011. I would like to thank Pietro Colangelo, Floriana Giannuzzi and Stefano Nicotri for stimulating discussions and many helpful comments.

%\newpage

%\bibliography{Mybib}
%\nocite{*}
%\newpage

\end{document}